\documentclass[aps,prl, twocolumn,floatfix,superscriptaddress]{revtex4}
\input epsf
\usepackage{epsfig}
\usepackage{amsmath,amssymb,latexsym}
\newcommand{\switchfonts}{\usefont{OT1}{cmr}{m}{it}}
\newcommand{\lie}{\switchfonts \mbox{\symbol{36}} \normalfont}

\newcommand{\sqrthb}{\sqrt{| \bar{h}| } }

\newcommand{\sigmat}{\Sigma_{t}}
\newcommand{\nablada}{ \bar{\nabla}_{a} }
\newcommand{\nabladb}{ \bar{\nabla}_{b} }

\begin{document}

\title{ Cosmological perturbation theory near de Sitter spacetime  }
\author{ B.~Losic }
\affiliation{ Department of Physics,
P-412, Avadh Bhatia Physics Laboratory
University of Alberta,
Edmonton, Alberta T6G 2J1
Canada}
\author{ W.G.~Unruh }
\affiliation{ Department of Physics \& Astronomy, 
University of British Columbia,
6224 Agricultural Road
Vancouver, B.C. V6T 1Z1
Canada }
\affiliation{ Canadian Institute for Advanced Research, Cosmology and Gravitation 
Program  }
\email{ blosic@phys.ualberta.ca   ;  unruh@physics.ubc.ca}

\date{April 21, 2008 }

\begin{abstract}
We present a gauge invariant argument that a nonlocal measure of second order metric and 
matter perturbations dominates that of linear fluctuations in its effect 
on the gravitational 
field in spacetimes close to the de Sitter solution.

\end{abstract}

\maketitle

{\it Introduction ---} It is well known in the mathematical physics community 
that linear cosmological perturbation theory about maximally symmetric, 
spatially closed, 
spacetimes has peculiar features. Linear perturbations about such special 
backgrounds must obey certain non-local identities (often discussed under 
the rubric of 
linearization stability \cite{F&M}, 
\cite{BrillDeser:1973}) which occur at second order in perturbation theory.

In this Letter we study a surprising consequence of these nonlocal identities 
for perturbations about a slowly-rolling (inflating) spacetime. We find 
that during 
'slow-roll' inflation a certain nonlocal measure of second order metric 
and matter perturbations generically dominates in its amplitude compared 
to that of the 
linear order perturbations, if these identities hold. This provides robust support for
the conclusions of one of our previous papers \cite{Losic:2005vg}, where we found that during slow-roll second order 
fluctuations grew large for a class of inflationary models. 
We conclude that is quite plausible that nonlinear, and probably nonperturbative, 
gravitational effects dominate near de Sitter spacetime (i.e. slow-roll) and therefore
linear perturbation theory likely fails in those situations. 


{\it Background model ---} Consider a FRW spacetime in comoving coordinates 
$(t,\vec{x})$ with scale factor $a(t)$, with signature (-1,1,1,1), and 
with a perfect fluid 
with energy density $\rho$ and pressure $p$. The equations of motion for 
the scale factor $a(t)$ are, according to the Einstein equations,
\begin{eqnarray}
\frac{ \ddot{a} }{a } &=& -\frac{ \kappa }{3} \left[ \rho (1 + 3 w) - \Lambda 
\right], \\
H^2 &=& \frac{  \kappa }{3} ( \rho + \Lambda)  - \frac{K}{a^2}, 
\end{eqnarray}
where $K \equiv \pm 1, 0$ is the constant spatial curvature of the $t = const$ 
slices, $H \equiv \partial_{t} ln( a )$ is the Hubble parameter, $\Lambda$ 
is a cosmological 
constant, $w \equiv \frac{p}{\rho}$, and $\kappa \equiv 8 \pi G$ in units 
where $c = 1$. 

There are in general six Killing vectors in FRW models, associated with 
either the $K = 0$ (flat) E(3) rotation group, the $K=-1$ (hyperbolic) 
SO(3,1) or $K=1$ (closed) 
O(4) groups. These three groups are maximal subgroups of the de Sitter 
group SO(4,1), which has ten parameters corresponding to the six FRW Killing 
vectors and four boost 
Killing vectors unique to de Sitter spacetime. The Lie derivative of the 
FRW metric along four vectors $B^{a}$ which have the same functional form 
of de Sitter boost Killing vectors in closed FRW coordinates can be easily 
calculated, using (1) and (2), to be
\begin{eqnarray}
\lie_{B} \bar{g}_{ab}^{(FRW)} = \lambda B^{i} \delta_{ia} \delta_{0b}
\end{eqnarray}
where the index $i$ above (and also $j,k$ in what follows) is spatial and 
$0$ refers to the `time` $t$, and where
\begin{eqnarray}
\lambda \equiv \frac{1}{2H} \left(  - \frac{\kappa \rho}{2} (1 + w) \right) 
\rightarrow 0
\end{eqnarray}
as one approaches de Sitter spacetime. Note that the four vectors $B^{a}$ are merely 
conformal isometries of the closed FRW spacetime.

{\it Nonlocal constraints ---} Consider the field equations for scalar 
matter and metric fluctuations for the above closed FRW solution, near 
de Sitter spacetime. 
Assuming that the background matter sector 
(i.e. $\rho, p$ in equations (1), (2)) is a minimally coupled ('potential 
dominated`) spatially homogeneous scalar field $\bar{\phi}$ with potential 
$V(\bar{\phi})$, a 
necessary requirement is that the fluctuations satisfy, order by order 
in perturbation theory, the initial value constraints on a constant time 
(spatially compact) 
hypersurface $\sigmat$
\begin{eqnarray}
{\cal{H}}_{\perp} &\equiv& {\cal{H}}_{\perp} ( h_{ij}, \pi^{ij}, \phi, 
\pi_{\phi}) = 0 \\
{\cal{H}}^{i} &\equiv& {\cal{H}}^{i}  (h_{ij}, \pi^{ij}, \phi, \pi_{\phi}) 
= 0
\end{eqnarray}
where equations (5) and (6) denote the usual 'Hamiltonian' and 'momentum' 
constraints respectively for the three metric $h_{ij}$, its conjugate momentum 
density 
$\pi^{ij}$, and the scalar field $\phi$ and its conjugate momentum $\pi_{\phi}$. These constraints must hold order by order in perturbation theory 
for a consistent power series approximation to exist, if one does, for 
a full solution to Einstein's equations. Also since the constraints hold at
each point in space, they must also hold when averaged with arbitrary
functions over space. 

Consider a projection (average)  of these constraints along an arbitrary 4-vector 
field $X$. Denoting this by $P(X)$, we write  

\begin{eqnarray}
P (X) &\equiv& \int_{\sigmat} \left( X_{\perp} {\cal{H}}_{\perp} + X_{\parallel}^{i} 
{\cal{H}}_{i} \right) d^3 x=0, 
\end{eqnarray}
where $X^{a} \equiv X_{\perp} \bar{n}^{a} + X_{\parallel}^{a} \bar{h}_{\ a}^{j}$is a four dimensional vector field where 
$n^{a}$ is the normal to $\sigmat$.  We wish to approximate 
$P(X)$ order by order in 
perturbation theory. 

Given a quantity $q$ we will designate the first order variation by $\delta q$ and the second order by $\delta^2 q$.  Furthermore we will designate the background quantities 
by an overbar $\bar{q}$. If we consider variations in $h_{ij}$ and $\pi^{ij}$ (along with $\phi$, $\pi_{\phi}$), we can  
calculate the corresponding classical variation in $P(X)$. We demand that the background quantities obey the full Einstein equations.  Using Hamilton's 
equations  for the background to {\it define} the time derivatives 
$\dot{\pi}^{ij}$, $\dot{h}_{ij}, \dot{\phi}, \dot{\pi}_{\phi}$ on a spatially compact $\sigmat$, one can show that 
\cite{Higuchi:1991tk}
\begin{eqnarray}
\delta P (X) &=& \int_{\sigmat} \left[ \left( \lie_{X} \bar{h}_{ij} \right)  
\delta \pi^{ij} - \left( L_{X} \bar{\pi}^{ij} \right) \delta h_{ij} \right. \\
\nonumber
				&& \left . + \left( \lie_{X} \bar{\phi} \right) \delta \pi_{\phi} - 
\left( \lie_{X} \bar{\pi}_{\phi} \right) \delta \phi \right] d^3 x,
\end{eqnarray}
where $\lie_{X} \bar{h}_{ij}$ is the spatial restriction of $\lie_{X} g_{ab}$ 
to the (spatially compact) hypersurface $\sigmat$:
\begin{eqnarray}
\lie_{X} \bar{h}_{ij} &=& \lie_{X_{\parallel}} \bar{h}_{ij} + X^{0} \dot{\bar{h}}_{ij} 
+ 2 \bar{N}_{(i} X^{0}_{|j)}, 
\end{eqnarray}
where $\bar{N}_{i} \equiv \bar{g}_{0i}$ is the `shift vector' (here and 
in what follows all barred quantities will be background quantities). 
The calculations are simplified if we take the {\bf background}
values of $\bar{N}_i=0$ and $\bar{N} \equiv -\bar{g}_{00} =1$. 

To give an idea how this is derived, consider the variation with respect to
$\delta\phi$. One of the terms in the above is
\begin{eqnarray}
\nonumber
&&\int_{\Sigma_{t}}  X^0\sqrt{|\bar{h}|}\left(h^{ij}\partial_i \bar\phi \partial_i \delta\phi+V'(\bar\phi)\delta \phi \right)d^3x\\
&=&\int_{\Sigma_{t}} X^0 \dot{\bar\pi} \delta\phi d^3x
\end{eqnarray}
because $\partial_i \bar{\phi}=0$ in the background. Similarly, 
\begin{eqnarray}
\int_{\Sigma_{t}} X^0 \frac{1}{\sqrt{|\bar{h}|}} \bar{\pi}_\phi \delta {\pi}_\phi d^3 x  =\int_{\Sigma_{t}} X^0\dot{\bar\phi}\delta\pi d^3x
\end{eqnarray}

Finally, using the metric equations of motion of the background spacetime,
we can show that 
\begin{eqnarray}
\nonumber
L_{X} \bar{\pi}^{ij} &=& X^{0} \dot{\bar{\pi}}^{ij} + \lie_{X_{\parallel}} 
\bar{\pi}^{ij} \\
&& + \sqrthb (\bar{D}^{i} \bar{D}^{j} - \bar{h}^{ij} \bar{D}_{k} \bar{D}^{k} ) X^{0},
\end{eqnarray}
where $\bar{D}_{i}$ is the induced covariant derivative on $\sigmat$.  
Putting equations (9) through (12) into equation (8) we finally obtain 
the general expression for the linearized projection of the initial value 
constraints, $\delta P(X)$. 

If we take the de Sitter limit, i.e. 
$V_{, \bar{\phi}}, \dot{\bar{\phi}} \rightarrow 0$, $\dot{H} \rightarrow 
0$, then $\lie_{X} \bar{h}_{ij} \rightarrow 0$ and $L_{X} \bar{\pi}^{ij} 
\rightarrow 0$ (and 
similarly for the matter fluctuations) yields $\delta P(X) \rightarrow 0$.  Thus, as is well known, the linearized projection of the constraint equations 
is identically zero along a Killing direction of the background spacetime 
provided the matter 
fluctuations obey the equations of motion (\cite{Higuchi:1991tk}, \cite{Moncrief:1976II}, 
\cite{Moncrief:1978te}, \cite{BrillDeser:1973}).  

The second order equations now have the form
\begin{eqnarray}
\nonumber
\delta^2 P(X) &=& \int_{\sigmat} \left[ \left( \lie_{X} \bar{h}_{ij} \right) \delta^2 \pi^{ij} 
					- \left( L_{X} \bar{\pi}^{ij} \right) \delta^2 h_{ij} \right. \\
\nonumber
            && \left . + \left( \lie_{X} \bar{\phi} \right) \delta^2 \pi_{\phi} - \left( \lie_{X} \bar{\pi}_{\phi} \right) \delta^2 \phi \right] d^3 x \\
&& + O(\delta q \delta q)
\end{eqnarray}
where the last term represents all of the terms quadratic in the first
order perturbations. 
This implies that in looking at the 
{\it second} order projection along the Killing vector(s), the terms linear
in the second order perturbations is zero, and the non-trivial quadratic
term must also be zero.  
This represents an additional constraint on the first order perturbations
which must be set to zero if the second order equations are to be
satisfied.

However as we discussed above it is clear that {\it near} a de Sitter spacetime 
one does not have exact {\it boost} symmetries.   If one projects the linearized 
constraint
denisities  ${\cal{H}}^{\alpha}$  along vectors $B^{a}$ which have a de 
Sitter boost {\it functional form} in closed FRW coordinates, as described 
above, then using 
equations (3) it follows from equations (8)-(12) that 
\begin{eqnarray}
\delta P(B) \propto \frac{\lambda}{H} \neq 0, 
\end{eqnarray} 
so that in the de Sitter limit, ( $\lambda \rightarrow 0$ ) the Killing 
identity is recovered. 

At second order in perturbation theory, we thus expect that $\delta^2 P(B)$ 
has {\it two} terms: one additional second order term multiplied by $\frac{\lambda}{H}$ 
and the 
quadratic piece. We compute $\delta^2 P(B)$ in the present Hamiltonian 
formalism and find
\begin{eqnarray}
 \delta^2 {P(B)_{\Sigma}}_{t} &=& \int_{\Sigma_{t}} B^{a} \delta^2 {\cal{H}}_{a} 
d^3x  \\
\nonumber
&=& \int_{\sigmat} \left[ \left( \lie_{B} \bar{h}_{ij} \right)  \delta^2 
\pi^{ij} - \left( L_{B} \bar{\pi}^{ij} \right) \delta^2 h_{ij} \right. 
\\
\nonumber
				&& \left . + \left( \lie_{B} \bar{\phi} \right) \delta^2 \pi_{\phi} 
- \left( \lie_{B} \bar{\pi}_{\phi} \right) \delta^2 \phi \right] d^3 x\\
\nonumber
&& +  \int_{\Sigma_{t}} \left[ \left( \lie_{B} \delta h_{ij} \right) \delta 
\pi^{ij} -  \left( L_{B} \delta \pi^{ij} \right) \delta h_{ij} \right 
. \\
\nonumber
	       && \left . + \delta \pi_{\phi} ( \lie_{B} \delta \phi )  - \delta 
\phi ( \lie_{B} \delta \pi_{\phi} ) \right] d^3 x, 
\end{eqnarray}
In the special case that $B^{a}$ {\it is} a Killing vector, i.e.  the background 
is closed vacuum de Sitter spacetime, it is clear that demanding the right 
hand side of (14) 
vanish implies a nontrivial and spatially nonlocal constraint on the linear 
initial values 
$(\delta h_{ij}, \delta \pi^{ij}; \delta \phi, \delta \pi_{\phi})$. In this case 
the nonlocal constraint, an integral over a density, is gauge invariant 
and preserved from slice to slice \cite{Moncrief:1978te}. 

{\it Slow-roll limit ---} It is apparent from equation (14) that there 
is an overall prefactor of $\frac{\lambda}{H}$ multiplying the second order 
terms compared to the 
final product term involving 
the linear fluctuations. Comparing the two groups of terms, second order 
($\equiv \delta^2 P_{S}(B)$) and quadratic in first order ($\equiv \delta^2 P_{Q}(B)$), 
we rework equation (15) by explcitly writing out the linear factor of $\frac{\lambda}{H}$ in $\delta^2 P_{S}$: 
\begin{eqnarray}
\delta^2 P(B) &=& \left( \frac{\lambda}{H} \right) \delta^2 P_{S}(B) [\delta^2 q_{i}] 
\\
\nonumber
						&& + \delta^2 P_{Q}(B) [ ( \delta q_{i} )^2, \delta q_{i} \delta 
q_{j} ],
\end{eqnarray}
where the entire set of second and linear order canonical variables is 
written as $\delta^2 q_{i}$ for the second order fluctuations, 
( $(\delta q_{i})^2, \delta q_{i} \delta q_{j} $) denotes the quadratic combinations 
of the first order fluctuations, and $\delta^2 P_{S}(B) \equiv \frac{\lambda}{H} \delta^2 \tilde{P}_{S}(B)$.
 Thus whenever the slow-roll approximation for the 
background holds, i.e. $\frac{\lambda}{H} \ll 1$,  we may approximately solve equation (16) for 
$\delta^2 P_{S}(B)  [ \delta^2 q_{i} ]$ to find 
\begin{eqnarray}
\delta^2 P_{S}(B)  [ \delta^2 q_{i} ] \approx - \frac{H}{\lambda} \delta^2 
P_{Q}(B) [(\delta q_{i})^2, \delta q_{i} \delta q_{j}  ]. 
\end{eqnarray} 

Thus, this combination of second order terms equals a large number times
some combination of the first order term. Assuming that the linear
fluctuations are not too small, this implies that at least this combination
of the second order fluctuation is larger than the first order
perturbations. 
This is  the main result of this paper: that {\it 
a nonlocal combination of second order metric and matter fluctuations will 
generically dominate in its effect on the projection of the gravitational 
constraints along $B^{a}$ compared to the linear terms }. Note that if 
$\delta q_{i} \ll \lambda / H$ the linearized fluctuations will not have the correct amplitude for seeding CMB fluctuations. 


{\it Gauge invariance ---} Equation (17), the main result of this paper, 
was derived without assuming a specific gauge choice. We now show that 
one 
cannot choose a second and/or linear order gauge such as to eliminate the 
factor of $\frac{H}{\lambda}$ in equation (17). 

Although $\delta P(B) \neq 0$ for any $\lambda \neq 0$, it is easy to show 
that the {\it background} projection $\bar{P}(B)$ actually vanishes identically 
for {\it any} value of $\lambda$, i.e. 
\begin{eqnarray}
\bar{P}(B) &=& 0
\end{eqnarray} 
for the background (closed FRW) constraints holding.
Thus $\delta^2 P(B)$ cannot depend on any purely second order infinitesimal 
coordinate transformation, just like any linear perturbation of a background 
constant is automatically gauge invariant to linear order. 

The most general remaining gauge transformation of equation (17) will induce 
an equation that can be written as 
\begin{eqnarray}
F_{S} ( 2 \lie_{\zeta} \delta q_{i}, {\lie_{\zeta}}^2 \bar{q}_{i}) &\approx& 
- \frac{H}{\lambda} F_{Q} ( 2 \lie_{\zeta} \delta q_{i}, {\lie_{\zeta}}^2 
\bar{q}_{i}),  
\end{eqnarray}
where $\zeta^{a}$ is an linearized (infinitesimal) coordinate transformation 
(so, e.g. $\delta^2 \pi^{ij} \rightarrow \delta ^2 \pi^{ij} + \lie^{2}_{\zeta} \bar{\pi}^{ij} + 2 \lie_{\zeta} \delta \pi^{ij}$) 
and $F_{S}$, $F_{Q}$ are the gauge terms coming from $\delta^2 P_{S}$ and $\delta^2 P_{Q}$ respectively. If one 
chooses 
\begin{eqnarray}
\nonumber
\zeta^{a} \equiv \left( \frac{\lambda}{H} \right)^{n} \tilde{\zeta}^{a}, n \in {\cal{Z}}^{+},
\end{eqnarray}
such that $n$ is the value required to eliminate the factor of $\frac{H}{\lambda}$ 
then one can rewrite (19) (by decomposing $F_{Q}$, $F_{P}$ into parts linear 
and quadratic 
in $\zeta^{a}$) as 
\begin{eqnarray}
&& \left( \frac{\lambda}{H} \right)^{n} \left[  {}^{(1)} f_{S} 
			(  2 \lie_{\tilde{\zeta}} \delta q_{i} ) +  \left( \frac{\lambda}{H} 
\right)^{n} {}^{(2)} f_{S} ( {\lie_{\tilde{\zeta}}}^2 \bar{q}_{i} ) \right] 
\\
\nonumber
&&  \approx - \frac{H}{\lambda} \left( \frac{\lambda}{H} \right)^{n} \left[ 
{}^{(1)} f_{Q} 
			(  2 \lie_{\tilde{\zeta}} \delta q_{i} ) +  \left( \frac{\lambda}{H} 
\right)^{n} {}^{(2)} f_{Q} ( {\lie_{\tilde{\zeta}}}^2 \bar{q}_{i} ) \right]
\end{eqnarray}
which clearly reduces to 
\begin{eqnarray}
{}^{(1)} f_{S} (  2 \lie_{\tilde{\zeta}} \delta q_{i} ) \approx - \frac{H}{\lambda} 
{}^{(1)} f_{Q} (  2 \lie_{\tilde{\zeta}} \delta q_{i} )
\end{eqnarray}
given that $\left( \frac{\lambda}{H} \right)^{n} \sim 0$, which is precisely 
of the form of equation (19). 

In summary, the form of equation (19) must persist given any first and 
second order gauge fixing in the perturbation theory, including
in particular the trivial choice $\zeta^{a} = 0$.  Another way of saying 
this is that the gauge dependence on both sides of equation (16) acts in 
such a way as to always 
preserve the form of equation (17). This is to be distinguished from the 
$\lambda = 0$ case, where the constraints (17) are {\it exactly} gauge 
invariant to second order.


{\it Quantum anomalies ---} The quadratic terms in equations (17) formally 
need to be regularized if we regard them as products of interacting quantum 
fields (see e.g., \cite{Hollands:2002ux}, \cite{Hollands:2001fb}, \cite{Hollands:2004yh}). 
Renormalization ambiguities could imply important quantum anomalies with respect 
to the imposition of 
second order conditions such as (17), in addition to any other reasonable 
conditions such as the conservation of stress-energy.

To begin with, one can show that there {\it will not} in general be anomalies 
associated with the simultaneous imposition of stress energy conservation 
and the 
equations of motion provided the background spacetime is slowly rolling. 
This is so because we can, in this very special case, specify the renormalization 
ambiguities 
(i.e. the nonuniqueness of a nonlinear monomomial (and its derivatives) 
in the fields) to absorb the considerably simplified slow-roll curvature 
counterterms. Specifically, for the case of the scalar field $\delta \phi$ it is known 
that the monomials 
$\Psi \equiv ( \delta \phi )^2$, $\Psi_{ab} \equiv \nablada \delta \phi \nabladb \delta \phi$ 
are unique up to the transformations \cite{Hollands:2004yh}
\begin{eqnarray}
\Psi &\rightarrow& \Psi + C \\
\Psi_{ab} &\rightarrow& \Psi_{ab} + C_{ab}
\end{eqnarray}
where $C$, $C_{ab}$ are quantities constructed from the metric $g_{ab}$, 
curvature, and derivatives of the curvature of the appropriate scaling 
dimension. For slow roll backgrounds, $C_{ab}$ and $C$ have a simple functional form.


Using this simplification one may show, just as we did for the case of 
pure de Sitter spacetime in \cite{Losic:2006ht}, that there are no {\it 
additional} 
anomalies associated with the imposition of the purely matter part of (17). 
This is so because all the anomaly terms are
proportional to integrals over $\Sigma_{t}$ of $B_{a} n^{a}$ (which even 
for all $\lambda \ge 0$ is spatially odd), which are identically zero.  
It turns out that 
{\it if} the remaining quadratic gravitational terms in (17) can be cast 
as quadratic scalar field terms 
(with some technical qualifications related to eliminating the homogeneous 
and dipole modes), where the scalar fields
represent polarizations of the transverse traceless excitations $\delta h_{ij}$, 
$\delta \pi^{ij}$ \cite{Ford:1977dj} and the lone scalar mode at linear 
order, then same 
logic goes through as for the scalar field case. One would then conclude, 
remarkably, that there are no additional quantum anomalies associated with 
the imposition of 
relations (17).  However, it should be strongly emphasized that in the 
absence of an explicit expression for the tensorial anomalies this is at 
best a plausible assumption ( see \cite{Losic:2007tu} ).

{\it Conclusions ---} In a previous publication \cite{Losic:2005vg} we have observed second order 
effects becoming large in some slow-roll models, however the 
present argument is demonstrably gauge invariant to second order and only 
essentially relies on the assumptions that the constraints are satisfied 
order by order in perturbation theory and that de Sitter spacetime has boost Killing vectors. 
Although our constraint analysis cannot answer the dynamical question of {\it when} (i.e. after 
how many e-foldings) 
these higher order effects can be expected to make a difference in typical 
slow-roll inflation or other models, our claim is that the worrisome higher 
order effects {\it do} 
unambiguously and rather generically enter with very minimal assumptions 
- they are really there.   We hope that by sidestepping the usual costly 
debate over whether or not 
higher order perturbative effects are just gauge effects or other artifacts 
of poorly controlled approximations, the present argument will serve as 
further motivation to probe 
higher order effects in cosmological perturbation theory near de Sitter 
spacetime.  

{\it Acknowledgements ---}
W.G.U. would like to thank the Canadian Institute for Advanced Research 
and NSERC for support. B.L. acknowledges YITP Workshop YITP-W-07-10 
(and YITP computer facilities) for hospitality and support, NSERC for support, 
and R. Brandenberger, E. Mottola, M. Sasaki, T. Tanaka, and R. Woodard
for many useful conversations. We also thank two anonymous referees for valuable
suggestions and criticism.

\end{document}